\documentclass[aps,prb,twocolumn]{revtex4-1}

\usepackage[pdftex]{graphics}
\usepackage{amssymb,amsmath}
\usepackage{nicefrac}
\usepackage{physics}
\usepackage{color}
\usepackage{braket}
\DeclareSymbolFont{boldoperators}{OT1}{cmr}{bx}{n}
\SetSymbolFont{boldoperators}{bold}{OT1}{cmr}{bx}{n}
\edef\bar{\unexpanded{\protect\mathaccentV{bar}}\number\symboldoperators16}

\begin{document}

\def\bra#1{\left<{#1}\right|}
\def\ket#1{\left|{#1}\right>}
\def\expval#1#2{\bra{#2} {#1} \ket{#2}}
\def\mapright#1{\smash{\mathop{\longrightarrow}\limits^{_{_{\phantom{X}}}{#1}_{_{\phantom{X}}}}}}

\title{Simple and accurate method for central spin problems}
\author{Lachlan P. Lindoy}
\affiliation{Department of Chemistry, University of Oxford, Physical and Theoretical Chemistry Laboratory, South Parks Road, Oxford, OX1 3QZ, UK}
\author{David E. Manolopoulos}
\affiliation{Department of Chemistry, University of Oxford, Physical and Theoretical Chemistry Laboratory, South Parks Road, Oxford, OX1 3QZ, UK}

\begin{abstract}
We describe a simple quantum mechanical method that can be used to obtain accurate numerical results over long time scales for the spin correlation tensor of an electron spin that is hyperfine coupled to a large number of nuclear spins. This method does not suffer from the statistical errors that accompany a Monte Carlo sampling of the exact eigenstates of the central spin Hamiltonian obtained from the algebraic Bethe ansatz, or from the growth of the  truncation error with time in the time-dependent density matrix renormalization group (t-DMRG) approach. As a result, it can be applied to larger central spin problems than the algebraic Bethe ansatz, and for longer times than the t-DMRG algorithm. It is therefore an ideal method to use to solve central spin problems, and we expect that it will also prove useful for a variety of related problems that arise in a number of different research fields.
\end{abstract}

\maketitle

The central spin Hamiltonian 
\begin{equation}
\hat{H} =  B \hat{S}_z+\sum_{j=1}^N a_j\hat{\bf S}\cdot\hat{\bf I}_j
\end{equation}
has been widely studied in the condensed matter physics literature because of its relevance to the hyperfine-induced decoherence of electron spins on quantum dots.\cite{Merkulov02,Khaetskii03,Erlingsson04,Braun05,Al-Hassanieh06,Chen07,Barnes12} It also arises as an important ingredient in the Hamiltonians that govern the dynamics of the electron spins in radical pairs of chemical and biochemical interest,\cite{Steiner89,Rodgers09,Hore12} and the spin dynamics of polaron pairs of interest to the organic semiconducting device community.\cite{Koopmans11,Wohlgenannt12,Ehrenfreund12} There is therefore considerable current interest in being able to calculate the exact quantum mechanical central ($\hat{\bf S}$) spin dynamics of the Hamiltonian in Eq.~(1).

In the context of an unpaired electron on a quantum dot, the first term in Eq.~(1) is the Zeeman interaction between the electron spin and an external magnetic field, and the second term is a sum of hyperfine interactions between the electron spin and the spins of the nuclei in the dot. The Zeeman interactions of the nuclear spins have been eliminated by defining $B=(g_{\rm e}-g_{\rm n})h$, where $g_{\rm e}$ and $g_{\rm n}$ are the electron and nuclear spin g-factors and $h$ is the magnetic field strength. Since $\hat{J}_z = \hat{S}_z+\sum_{j=1}^N \hat{I}_{jz}$ is a constant of the motion, the difference $g_{\rm n}h\hat{J}_{z}$ between Eq.~(1) and the Hamiltonian that includes nuclear Zeeman interactions is a trivial energy shift within each $\hat{J}_z$ symmetry block.\cite{Faribault13b} The dipolar interactions between the nuclear spins in the dot have been neglected, but since they are much weaker than the terms that have been retained in Eq.~(1) they will only affect the electron spin dynamics at very long times.

The difficulty of the problem lies in the fact that the size of the Hilbert space increases exponentially with $N$. Assuming for simplicity that all of the nuclear spins have $I=1/2$, the dimension of the Hilbert space is $2^{N+1}$. This makes solving the problem by numerical diagonalization  impractical for $N\gtrsim 20$. However, quantum dots typically have many more than 20 nuclear spins with non-negligible hyperfine interactions, and the same is true of many interesting radicals and polarons.

In the condensed matter physics literature, several highly sophisticated techniques have been developed to overcome the exponential scaling. Perhaps the most elegant of these exploits the fact that the Hamiltonian in Eq.~(1) has the form of a Gaudin magnet in an external magnetic field, which is well known to be an integrable problem.\cite{Gaudin76,Gaudin83} In other words, there are $N$ constants of the motion in addition to $\hat{H}$, of which $\hat{J}_z$ is just one. The eigenvalues and eigenstates of all $N+1$ conserved operators, and therefore of $\hat{H}$ itself, can be found by solving a system of $N+1$ coupled algebraic Bethe equations.\cite{Faribault13a,Faribault13b} The solution of these equations is obvious when $1/B\to 0$, and this solution can be continued numerically to larger $1/B$.\cite{Faribault13b} But once this has been done, one is still faced with the problem of summing over all of the eigenstates of $\hat{H}$ to calculate the dynamics of the central electron spin, with a numerical effort that again scales exponentially with $N$. Faribault and Schuricht have suggested a Monte Carlo sampling of the eigenstates to alleviate this problem, and demonstrated that this implementation of the algebraic Bethe ansatz is capable of solving central spin problems with up to $N=48$ nuclear spins.\cite{Faribault13b} However, the statistical errors from the Monte Carlo sampling are quite noticeable in their results, and one would expect these errors to become even larger for larger $N$.

A contrasting approach has been to adapt the time-dependent density matrix renormalization group (t-DMRG) method to the topology of the central spin problem.\cite{Stanek13,Stanek14} Since the t-DMRG algorithm is best suited to a one-dimensional chain of spins with nearest-neighbour interactions, this requires some ingenuity. It can however be done, as Uhrig and co-workers have shown in a pair of papers that include well converged short-time results for a number of central spin problems with up to 999 hyperfine-coupled nuclear spins.\cite{Stanek13,Stanek14} The trouble with this method is that it is limited to short time scales. From Eq.~(1), the coupling between any two nuclear spins via the electron spin involves only two successive Hamiltonian interactions. This implies that the entanglement of the exact time-evolved wavefunction will grow rapidly with time, and that it will become increasingly difficult to represent it with the tensor product form that is assumed in t-DMRG. Hence the truncation error in this algorithm is bound to grow with time. Uhrig and co-workers have shown that this error becomes unacceptable beyond 40$\tau$ for a problem with $N=99$ nuclear spins, where
\begin{equation}
\tau = \left(\sum_{j=1}^N a_j^2\right)^{-{1\over 2}}
\end{equation}
is the characteristic time scale of the electron spin precession in the nuclear hyperfine field. The t-DMRG truncation error is also expected to grow with $N$.

Motivated by these issues with existing algorithms, we shall now present an entirely different approach to the problem. Rather than attempting to calculate the central spin dynamics of the Hamiltonian in Eq.~(1), we shall instead construct a sequence of simpler Hamiltonians $\hat{H}_{M}$ for $M=1,2,\ldots$ in such a way that their central spin dynamics converges to that of the original Hamiltonian with increasing $M$.

The Hamiltonians $\hat{H}_M$ we shall consider are
\begin{equation}
\hat{H}_M = B\hat{S}_z+\sum_{j=1}^M A_j\sum_{i=1}^{N_j} \hat{\bf S}\cdot\hat{\bf I}_{ij},
\end{equation}
in which the modified hyperfine coupling constants $A_j$ and the numbers $N_j$ of equivalent nuclear spins in each of the $M$ blocks are chosen to ensure that the first $M+1$ moments of the modified hyperfine distribution coincide with those of the original distribution:
\begin{equation}
\mu_k = \sum_{j=1}^N a_j^k = \sum_{j=1}^M N_j {A}_j^k \hbox{   for   } k=0,1,\ldots,M.
\end{equation}
In order to obtain these Hamiltonians, we first use a discrete procedure of Stieltjes\cite{NR} to construct a Gaussian quadrature rule with non-integer weights $W_j$ and nodes $\bar{A}_j$ such that 
\begin{equation}
\mu_k = \sum_{j=1}^N a_j^k = \sum_{j=1}^M W_j\bar{A}_j^k \hbox{   for   } k=0,1,\ldots,2M-1.
\end{equation}
We then search through all floor and ceiling possibilities for the integers $N_j = \left\lfloor{W_j}\right\rfloor$ or $\left\lceil{W_j}\right\rceil$ in Eq.~(4) that are consistent with the $k=0$ moment constraint
\begin{equation}
\sum_{j=1}^M N_j = N,
\end{equation}
and use Newton's method to solve the remaining moment equations in Eq.~(4) for the $M$ unknowns $\left\{A_j\right\}$, starting from the initial guess $\left\{A_j\right\}=\left\{\bar{A}_j\right\}$. Finally, we select from among the $< 2^M$ possible solutions for the set of integers $\left\{N_j\right\}$ the one that minimises the error in $\mu_{M+1}$.

The reason for choosing simplified Hamiltonians $\hat{H}_M$ of the form in Eq.~(3) is that the presence of equivalent nuclei dramatically simplifies the spin dynamics calculation. For example, a calculation with $M=1$ and $N_1=100$ equivalent spin-1/2 nuclei can be reduced by symmetry to 51 separate calculations, each of which involves a single resultant spin $\hat{\bf I}_1=\sum_{i=1}^{N_1} \hat{\bf I}_{i1}$ with $I_1$ between 0 and 50. The Hilbert space of the largest of these calculations has dimension $2\times 101=202$, whereas the dimension of the full Hilbert space is $2^{101}$. A calculation with $M=2$ and $N_1=N_2=50$ can be reduced in the same way to $26\times 26=676$ separate calculations with $I_1$ and $I_2$ between 0 and 25. The Hilbert space of the largest of these calculations has dimension $2\times 51\times 51=5,202$. And so on. For a given $N$, the effort increases with $M$, but it remains many orders of magnitude smaller than that of the full calculation for all $M\ll N$.

The choice of model Hamiltonians $\hat{H}_M$ that have hyperfine distributions with the same first moments as those of the exact Hamiltonian $\hat{H}$ is physically motivated. The early semiclassical theory of Schulten and Wolynes\cite{Schulten78} only involves the second moment $\mu_2$ of the hyperfine distribution, and yet it becomes exact for the central spin dynamics in the limit as $N\to\infty$.  By constructing our model Hamiltonians on the basis of Eq.~(4), we are therefore guaranteed to obtain the correct result in the large $N$ limit for any $M\ge 2$. And since Eq.~(3) reduces to Eq.~(1) when $M=N$ (with $A_j=a_j$ and $N_j=1$ for all $j$), we are also guaranteed to obtain the correct result in the small $N$ limit in which a calculation based on Eq.~(1) is feasible. The only remaining question, therefore, is how well this method actually works for the intermediate values of $N$ that are of interest in physical applications.

In order to answer this question, we have used the method to calculate the infinite temperature central spin correlation tensors
\begin{equation}
R_{\alpha\beta}(t) = {1\over Z}{\rm tr}\left[\hat{S}_{\alpha}(0)\hat{S}_{\beta}(t)\right],
\end{equation}
with $Z=2^{N+1}$, for two model problems that have been studied previously using the t-DMRG and algebraic Bethe ansatz methods. The first of these has a uniform distribution of hyperfine coupling constants\cite{Stanek13}
\begin{equation}
a_j\tau = \sqrt{6N\over 2N^2+3N+1}{{N+1-j}\over N},
\end{equation}
and the second has an exponential distribution that arises from the Fermi contact interactions in a two-dimensional quantum dot with a Gaussian electronic wavefunction\cite{Faribault13b}
\begin{equation}
a_j\tau = \sqrt{{1-e^{-2/(N_0-1)}}\over 1-e^{-2N/(N_0-1)}}e^{-(j-1)/(N_0-1)},
\end{equation}
where $N_0$ is the number of nuclear spins within one Bohr radius of the Gaussian wavefunction and $\tau\equiv \sqrt{1/\mu_2}$ is the timescale in Eq.~(2).

In what follows, we shall refer to these two models as Model I and Model II, respectively. Uhrig and co-workers used the t-DMRG method to study the low-field ($B=0$) limit of Model I for $N=49$, 99, and 999,\cite{Stanek13,Stanek14} and Faribault and Schuricht used the algebraic Bethe ansatz to study the low field limit of Model II with $N=48$ and $N_0=24$ and 36.\cite{Faribault13b} These are the model problems we shall consider. Since both have $B=0$, we have that $R_{xx}(t)=R_{yy}(t)=R_{zz}(t)$ and $R_{xy}(t)=R_{yz}(t)=R_{zx}(t)=0$. It therefore suffices to consider just $R_{zz}(t)$, from which all other properties of the central spin dynamics can be recovered.\cite{footnote1}

\begin{figure}[t]
\centering
\resizebox{0.9\columnwidth}{!} {\includegraphics{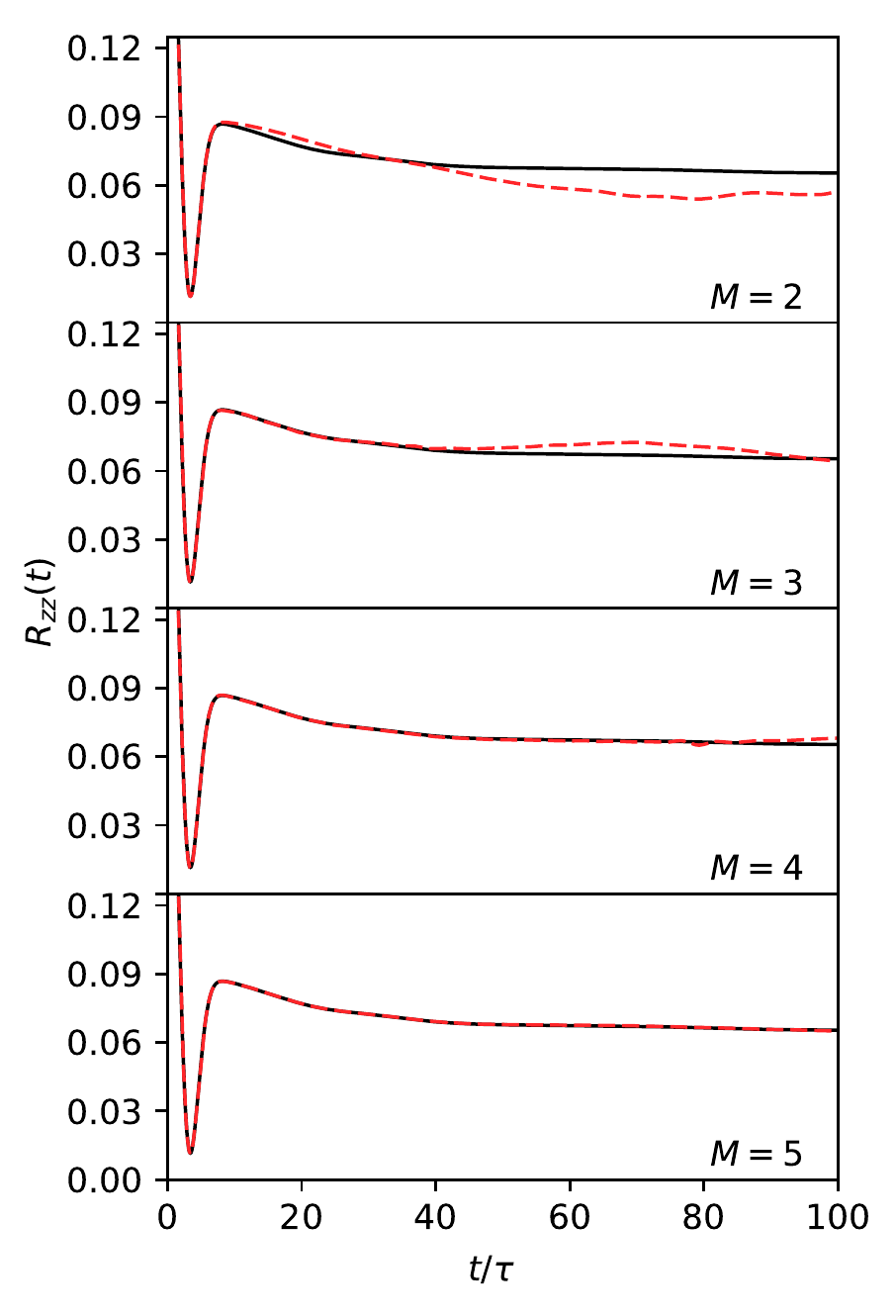}}
\caption{Convergence of $R_{zz}(t)$ for Model I with $N=49$, as a function of $M$. The solid black curve in each panel shows the fully-converged result obtained with $M=7$, and the dashed red curve shows the result obtained with the specified value of $M$.}
\end{figure}

We have calculated $R_{zz}(t)$ for both of these models using the method outlined above. Rather than numerically diagonalising the matrix representations of the Hamiltonians $\hat{H}_M$ in Eq.~(3), we found it more efficient to use a symplectic time-dependent wavepacket propagation algorithm.\cite{Gray96} The traces in Eq.~(7) were evaluated deterministically for each symmetry block of $\hat{H}_M$ containing less than 1000 states, and stochastically using a recently-developed spin coherent state algorithm\cite{Lewis16} for the larger symmetry blocks. Note that this is quite different from stochastically sampling the eigenstates of the full Hamiltonian $\hat{H}$ as was done by Faribault and Schuricht.\cite{Faribault13b} Our statistical errors were found to be negligible with just 1000 Monte Carlo samples of the initial nuclear spin coherent states in each of the larger symmetry blocks.

\begin{figure}[t]
\centering
\resizebox{0.9\columnwidth}{!} {\includegraphics{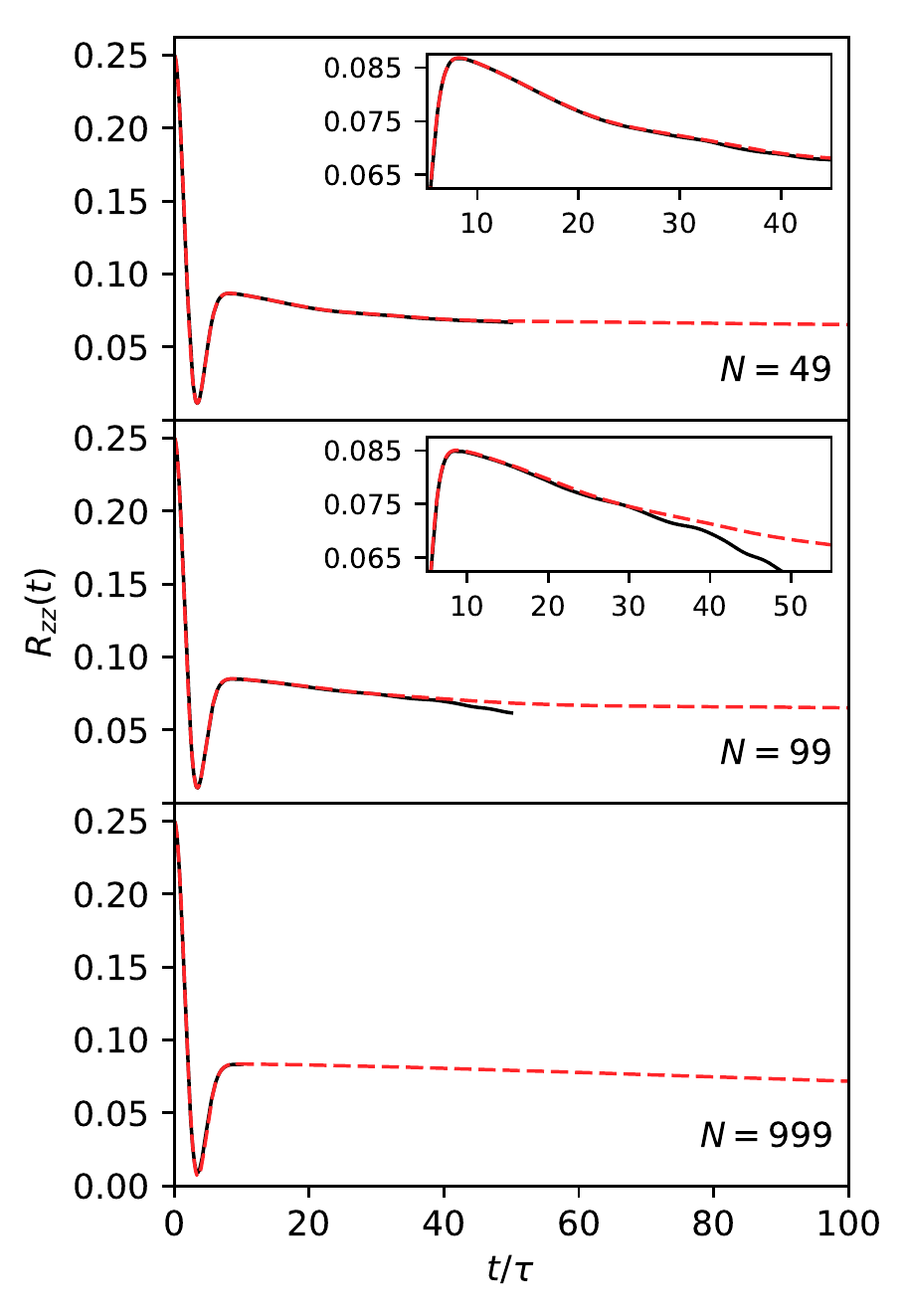}}
\caption{Comparison of the present results for Model I (dashed red curves) with the t-DMRG results of Uhrig and co-workers\cite{Stanek13,Stanek14} (solid black curves),  for $N=49$, 99, and 999.}
\end{figure}

Fig.~1 shows the convergence of the correlation function $R_{zz}(t)$ obtained from the present method with increasing $M$, for Model I with $N=49$ nuclear spins. One sees that the method remains accurate for longer times as $M$ increases, and that it is fully converged over the time interval considered (up to $t=100\tau$) by the time $M=5$. Similar convergence plots for  $N=99$ and 999 are given in the supplementary material. Full convergence out to $100\tau$ is obtained with $M=4$ for $N=99$, and with just $M=3$ for $N=999$. Thus the method becomes increasingly efficient with increasing $N$. Presumably this is because $N=999$ is approaching the large-$N$ limit in which the central spin dynamics is determined entirely by the second moment of the hyperfine distribution.

\begin{figure}[t]
\centering
\resizebox{0.9\columnwidth}{!} {\includegraphics{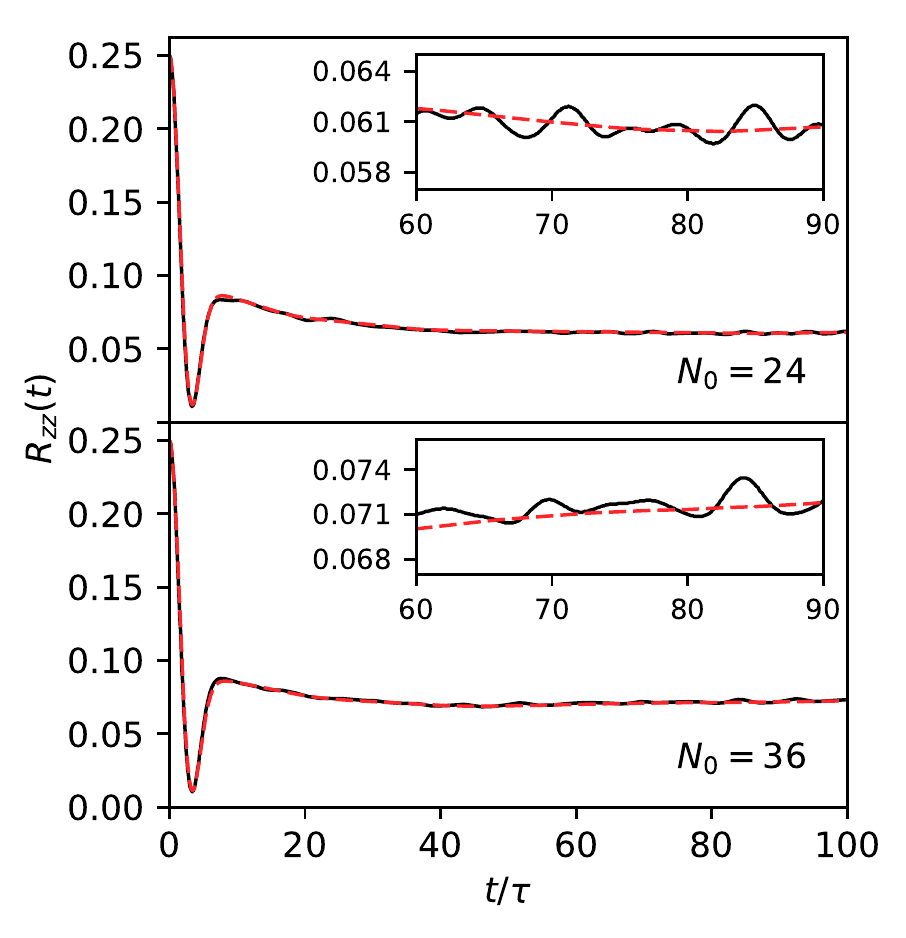}}
\caption{Comparison of the present results for Model II (dashed red curves) with the algebraic Bethe ansatz results of Faribault and Schuricht\cite{Faribault13b} (solid black curves), for $N=48$.}
\end{figure}

Fig.~2 compares our converged results for Model I with the t-DMRG results of Uhrig and co-workers.\cite{Stanek13,Stanek14} The good agreement over the timescales for which the t-DMRG results are available confirms that the present method converges on the correct quantum mechanical central spin dynamics of this model. It also avoids the growth of the truncation error with time in the t-DMRG algorithm, which is especially apparent in the inset of the $N=99$ panel in the figure. The t-DMRG results for $N=999$ are only available out to $t=10\tau$ because the truncation error increases with increasing $N$.

\begin{figure}[t]
\centering
\resizebox{0.9\columnwidth}{!} {\includegraphics{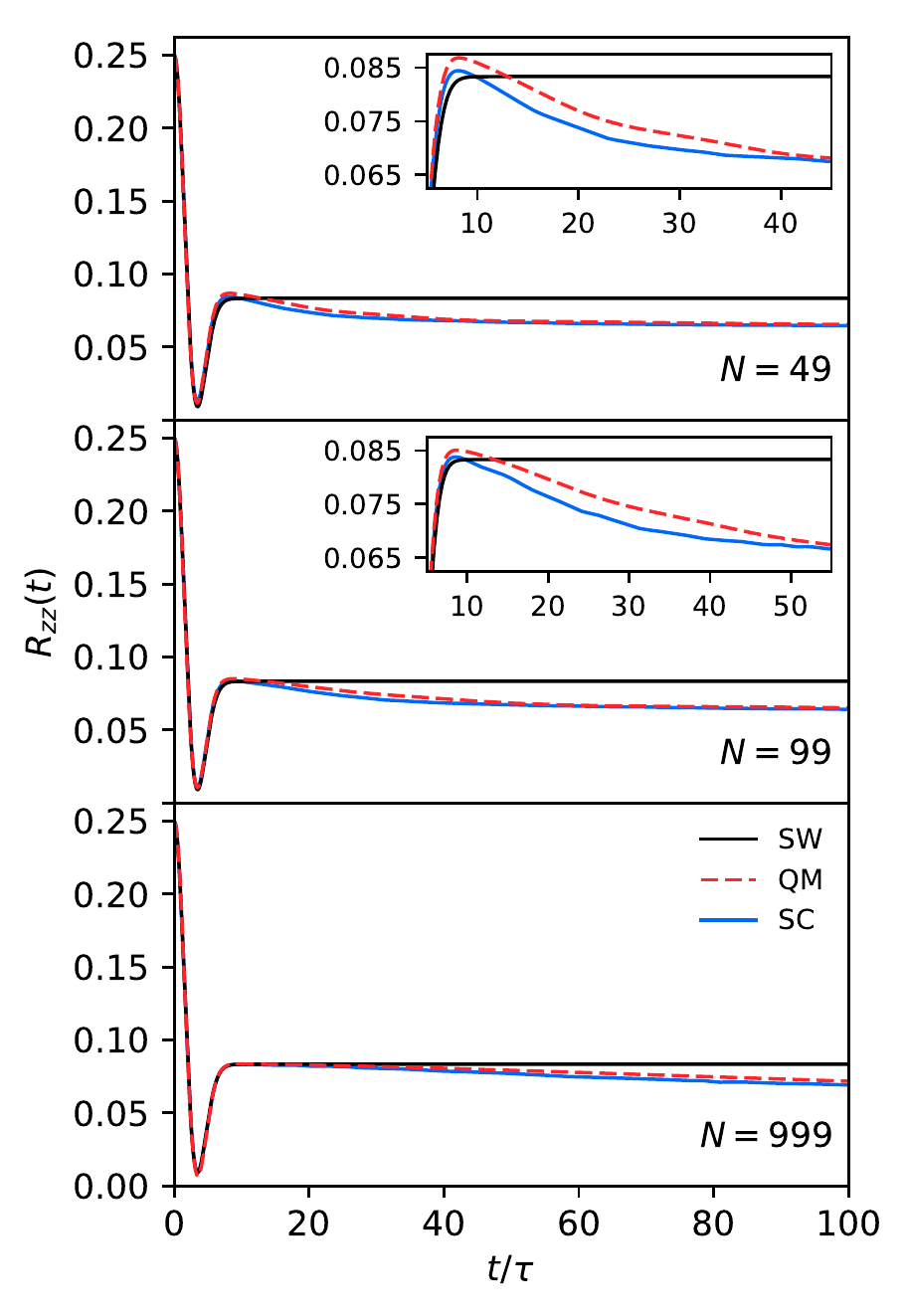}}
\caption{Comparison of the present quantum mechanical (QM) results for Model I with the results of the early semiclassical theory of Schulten and Wolynes\cite{Schulten78} (SW) and the more recent semiclassical theory of Manolopoulos and Hore\cite{Manolopoulos13} (SC). The long-time behaviours of these correlation functions are discussed in the supplementary material.}
\end{figure}

\begin{figure}[t]
\resizebox{0.9\columnwidth}{!} {\includegraphics{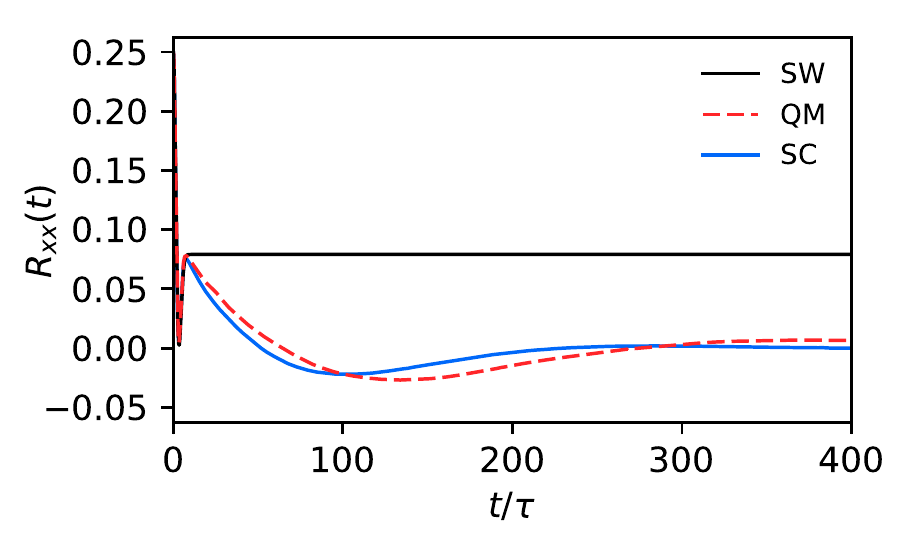}}
\caption{Comparison of SW, QM, and SC results for $R_{xx}(t)$ for Model I with $N=49$ in a magnetic field of strength $B=1/4\tau$. Note that the time scale available from a t-DMRG calculation ($t<50\tau$) would not be enough to predict the long-time behaviour of this correlation function, which is discussed further in the supplementary material.}
\end{figure}

The convergence tests we have performed for Model II (with $N=48$ nuclear spins) are provided in the supplementary material. $M=5$ was again found to be sufficient to converge $R_{zz}(t)$ out to $t=100\tau$, in much the same way as for Model I with $N=49$ nuclear spins. Fig.~3 compares the resulting correlation functions with those obtained by Faribault and Schuricht using the algebraic Bethe ansatz.\cite{Faribault13b} The agreement between the two sets of calculations is again excellent, and confirms that both methods are giving the correct correlation functions for this model. The only difference between the two sets of results is that our correlation functions are smoother than those of Faribault and Schuricht, which have noticeable stochastic errors associated with the incomplete Monte Carlo sampling of the eigenstates of $\hat{H}$ in their method. As we have already mentioned, these stochastic errors in the algebraic Bethe ansatz method are expected to become even more pronounced for larger $N$.

Finally, let us return to Model I and the question of how close $N=999$ is to the large $N$ limit in which the early semiclassical theory of Schulten and Wolynes\cite{Schulten78} is expected to become exact. Fig.~4 compares the present quantum mechanical results for this model with those given by their theory,  and also with those of an improved semiclassical theory suggested by Manolopoulos and Hore.\cite{Manolopoulos13,footnote2} 

The quantum mechanical results in Fig.~4 are sufficiently well converged, and for sufficiently long times, to provide a stringent test of these semiclassical approximations. One sees from the figure that, although both semiclassical theories are qualitatively reasonable, neither is quantitatively accurate, even for $N=999$. The Schulten-Wolynes theory misses the long-time decay of the central spin correlation function, and the improved semiclassical theory predicts too rapid a long-time decay. This is even more apparent in Fig.~5, which shows $R_{xx}(t)$ over a longer time scale for Model I with $N=49$ in a magnetic field of strength $B=1/4\tau$.

In view of this, it would be interesting to apply the present method to a variety of problems that have previously only been studied semiclassically. One such problem is the spin dynamics of a photoexcited carotenoid-porphyrin-fullerene radical pair that has been shown to be sensitive to an Earth strength magnetic field.\cite{Maeda08} Since this field is so weak ($\sim 50\ \mu$T), and the experiment that was used to detect it measured a tiny field-on minus field-off difference signal,\cite{Maeda08} it is conceivable that the semiclassical calculations that have been used to study the problem\cite{Lewis14} might not have been sufficiently accurate. The carotenoid radical in the pair contains 45 protons with significant hyperfine interactions, which have so far precluded an exact quantum mechanical calculation. The results we have presented here show that the present method could easily handle this problem, and many other interesting spin dynamics problems as well.

\begin{acknowledgements}
We are grateful to Pranav Singh for a helpful comment on the first draft of this manuscript. Lachlan Lindoy is supported by a Perkin Research Studentship from Magdalen College, Oxford, an Eleanor Sophia Wood Postgraduate Research Travelling Scholarship from the University of Sydney, and by a James Fairfax Oxford Australia Scholarship. The authors would like to acknowledge the use of the University of Oxford Advanced Research Computing (ARC) facility in carrying out this work. See http://dx.doi.org/10.5281/zenodo.22558.
\end{acknowledgements}


\begin{thebibliography}{100}
\bibitem{footnote1}
For example, the observable considered by Faribault and Schuricht in Ref.~14 is ${\rm Re}\left[\left<S_+(t)\right>\right]/\left<S_+(0)\right>$ with $\rho(0)=(2S_x+1)/Z$. It is easy to show that this is the same as $4R_{xx}(t)$, which coincides with $4R_{zz}(t)$ when $B=0$.
\end{thebibliography}
\end{document}


\def\bra#1{\left<{#1}\right|}
\def\ket#1{\left|{#1}\right>}
\def\expval#1#2{\bra{#2} {#1} \ket{#2}}
\def\mapright#1{\smash{\mathop{\longrightarrow}\limits^{_{_{\phantom{X}}}{#1}_{_{\phantom{X}}}}}}

\title{A simple and accurate method for central spin problems:\\ Supplementary material}
\author{Lachlan P. Lindoy}
\affiliation{Department of Chemistry, University of Oxford, Physical and Theoretical Chemistry Laboratory, South Parks Road, Oxford, OX1 3QZ, UK}
\author{David E. Manolopoulos}
\affiliation{Department of Chemistry, University of Oxford, Physical and Theoretical Chemistry Laboratory, South Parks Road, Oxford, OX1 3QZ, UK}

\begin{abstract}
This supplementary material contains four sections. The first provides the computer program that we used to generate the simplified Hamiltonians $\hat{H}_M$ in Eq.~(3) of the text, along with example input and output files. This information is provided so as to make our results reproducible, and to enable others to apply our method to other spin dynamics problems. The second contains more details on how to exploit the presence of equivalent spins in spin dynamics calculations, and the third contains plots showing the convergence of our method with increasing $M$ for Model I with $N=99$ and 999, and for Model II with $N_0=24$ and 36. The final section discusses semi-log plots ($R_{\alpha\alpha}(t)$ vs $\log t$) of the data in Figs.~4 and 5 of the manuscript, which provide a clearer picture of the long-time behaviour of the various (Schulten-Wolynes, semiclassical, and quantum mechanical) central spin correlation functions than the linear plots given in the manuscript.
\end{abstract}
\maketitle

\section{How to construct the Hamiltonians $\hat{H}_M$}

The text of the paper has simply outlined how we have constructed the simplified Hamiltonians $\hat{H}_M$ in Eq.~(3). Here we provide the computer code that we have actually used to do this, so that others can both reproduce our results and apply the same method to other spin dynamics problems. 

The following fortran program reads in 3 input parameters, $N_0$, $N$, and $M$. It then constructs $\hat{H}_M$, and outputs the optimised hyperfine coupling constants $\left\{A_j\right\}_1^M$ and numbers $\left\{N_j\right\}_1^M$ of equivalent nuclear spins in each of its symmetry blocks. It also outputs the percentage errors in the first $M+1$ moments of the hyperfine distribution that result from replacing the original Hamiltonian with the simplified Hamiltonian.

The results for the first central spin problem (Model I) considered in the text were obtained by inputting $N_0=0$, $N=49$, 99, and 999, and $M=5$, 4, and 3, respectively. The results for the second problem (Model II) were obtained by inputting $N_0=24$ and $36$, $N=48$, and $M=5$. The convergence tests we shall present below were obtained using other values of $M$ between 2 and 7.

\begin{verbatim}

      program moments
      implicit double precision (a-h,o-z)
c
c     ------------------------------------------------------------------ 
c     This program constructs a simplified central spin Hamiltonain 
c     with M sets of equivalent nuclei from a given Hamiltonian with 
c     N inequivalent nuclei.
c     ------------------------------------------------------------------ 
c
      allocatable :: a(:),abar(:),nbar(:)
c
c     Setup:
c
      read (5,*) n0,n,m
      allocate (a(n),abar(m),nbar(m))
      if (n0 .eq. 0) then! uniform hyperfine distribution 
         fac = sqrt((6.0d0*n)/(2.0d0*n*n+3.0d0*n+1.0d0))
         do j = 1,n
            a(j) = fac*(n-(j-1.0d0))/n
         enddo
      else if (n0 .gt. 1) then! exponential hyperfine distribution 
         top = 1.d0-exp(-2.d0/(n0-1.d0))
         btm = 1.d0-exp(-(2.d0*n)/(n0-1.d0))
         fac = sqrt(top/btm)
         do j = 1,n
            a(j) = fac*exp(-(j-1.d0)/(n0-1.d0))
         enddo
      else
         stop 'moments 1'
      endif
c
c     Calculation:
c
      call shrink (a,n,abar,nbar,m)
c
c     Output:
c
      write (6,600) n0,n,m
 600  format(1x,' N0 = ',i5,'   N = ',i5,'   M = ',i5/1x,/1x,
     + '     j      N_j                 A_j'/1x)
      ntot = 0
      do j = 1,m
         write (6,601) j,nbar(j),abar(j)
 601     format(1x,i6,i9,f20.12)
         ntot = ntot+nbar(j)
      enddo
      if (ntot .ne. n) stop 'moments 2'
      write (6,602)
 602  format(/1x,'     k              % error in mu_k'/1x)
      do k = 1,m+1
         exact = 0.d0
         do i = 1,n
            exact = exact+a(i)**k
         enddo
         approx = 0.d0
         do j = 1,m
            approx = approx+nbar(j)*abar(j)**k
         enddo
         error = 100.d0*abs(exact-approx)/abs(exact)
         write (6,603) k,error
 603     format(1x,i6,f29.6)
         if (k .eq. m) print*
      enddo 
      deallocate (a,abar,nbar)
      stop
      end

      subroutine shrink (a,n,abar,nbar,m)
      implicit double precision (a-h,o-z)
c
c     ------------------------------------------------------------------ 
c     Optimally approximates N inequivalent nuclei with hyperfine
c     coupling constants a(j) by M < N sets of equivalent nuclei
c     with hyperfine constants abar(j), with nbar(j) nuclei in set j.
c     ------------------------------------------------------------------ 
c
c     We use integer*8 so that we can deal with all M < 64:
c
      integer*8 ib,nset
      dimension a(n),as(n),awbar(m),abar(m),nbar(m),nfloor(m)
      dimension atemp(m),ntemp(m)
      dimension w(n),wbar(m)
c
c     It only makes any sense to call this subroutine with M < N:
c
      if (m .gt. n) stop 'M > N in shrink?'
      if (m .eq. n) stop 'M = N in shrink?'
c
c     Optimum solution with non-integer weights:
c
      do i = 1,n
         w(i) = 1.d0
      enddo
      call qrule (n,w,a,m,wbar,awbar)
      nftot = 0
      do j = 1,m
         nfloor(j) = int(wbar(j))
         nftot = nftot+nfloor(j)
      enddo
      ndiff = n-nftot
c
c     Optimum solution with integer weights:
c
      dmom = 1.d10
      do ib=0,LSHIFT(1,m)-1
         call count_set_bits(ib,nset)
         if(nset .eq. ndiff) then
            do i=1,M
                nind = RSHIFT(IAND(ib,LSHIFT(1,i-1)),i-1)
                ntemp(i) = nfloor(i)+nind 
                atemp(i) = awbar(i)
            enddo
            call newton (awbar,wbar,atemp,ntemp,m,icode)
            if (icode.eq.0) then
                emom = 0.0d0
                do k = 1,m+1
                   exact = 0.0d0
                   do i = 1,n
                      exact = exact+a(i)**k
                   enddo
                   approx = 0.0d0
                   do j = 1,m
                      approx = approx+ntemp(j)*atemp(j)**k
                   enddo
                   emom = emom+abs(approx-exact)
                enddo 
                if (emom.lt.dmom) then
                    dmom = emom
                    do i = 1,m
                        nbar(i) = ntemp(i)
                        abar(i) = atemp(i)
                    enddo
                endif
            endif
         endif
      enddo
      return
      end

      subroutine count_set_bits (i,nset)
      implicit none
c
c     ------------------------------------------------------------------ 
c     Determines the total number of non-zero bits in a 64 bit integer 
c     using a 64 bit implementation of the Hamming Weight algorithm.
c     ------------------------------------------------------------------ 
c
      integer*8 i,nset,v
      integer*8 m1,m2,m3,m4
      DATA m1 /Z'5555555555555555'/, m2/Z'3333333333333333'/
      DATA m3 /Z'f0f0f0f0f0f0f0f'/, m4/Z'101010101010101'/
c
      v = i
      v = v-IAND(RSHIFT(v,1),m1)
      v = IAND(V,m2)+IAND(RSHIFT(V,2),m2)
      v = IAND(v+RSHIFT(v,4),m3)
      nset = RSHIFT(v*m4,(SIZEOF(nset)-1)*8)
      return
      end

      subroutine newton (awbar,wbar,abar,nbar,m,icode)
      implicit double precision (a-h,o-z)
c
c     ------------------------------------------------------------------ 
c     Uses Newton's method to solve M non-linear moment equations in M
c     unknowns (the hyperfine constants of M sets of equivalent nuclei),
c     starting from an appropriate initial guess.
c     ------------------------------------------------------------------ 
c
      parameter (maxit = 30)
      dimension awbar(m),wbar(m),abar(m),nbar(m)
      dimension f(m),g(m,m),indx(m)
c
      do iter = 1,maxit
         do k = 1,m
            f(k) = 0.d0
            do j = 1,m
               f(k) = f(k)+nbar(j)*abar(j)**k-wbar(j)*awbar(j)**k
               g(k,j) = k*nbar(j)*abar(j)**(k-1)
            enddo
         enddo
         call rgfac (g,m,m,indx,ierr)
         if (ierr .ne. 0) then              
            icode = -1
            return
         endif
         call rgsol (g,m,m,indx,f)
         absa = 0.d0
         absd = 0.d0
         do j = 1,m
            absa = absa+abs(abar(j))
            absd = absd+abs(f(j))
            g(j,1) = abar(j)
            abar(j) = abar(j)-f(j)
         enddo
         da = absd/absa
         if (da .gt. 1.d0) then
            icode = 1
            return 
         endif
         if (1.d0+da .eq. 1.d0) then
            icode = 0
            return
         endif
         if (iter.gt.1 .and. da.ge.dap) then
            do j = 1,m
               abar(j) = g(j,1)
            enddo      
            icode = 0
            return
         endif
         dap = da
      enddo
      icode = 2
      return
      end

      subroutine qrule (np,wp,xp,n,w,x)
      implicit double precision (a-h,o-z)
c
c     ----------------------------------------------------------------- 
c     Uses a discrete Stieltjes procedure to construct an n-point 
c     contracted quadrature rule from an np-point primitive quadrature 
c     rule with the same (non-negative) weight function.
c     ----------------------------------------------------------------- 
c      
      dimension wp(np),xp(np),w(n),x(n)
      dimension p(np,n),q(np),v(n)
c
      do i = 1,np
         q(i) = sqrt(wp(i))
      enddo
      do k = 1,n
         qq = 0.d0
         do i = 1,np
            qq = qq+q(i)**2
         enddo
         qq = sqrt(qq)
         if (qq .eq. 0.d0) stop 'qrule 1'
         do i = 1,np
            p(i,k) = q(i)/qq
            q(i) = xp(i)*p(i,k)
         enddo
         do j = k,1,-1
            pq = 0.d0
            do i = 1,np
               pq = pq+p(i,j)*q(i)
            enddo
            if (j .eq. k) then
               v(k) = 0.d0
               w(k) = qq
               x(k) = pq
            endif
            do i = 1,np
               q(i) = q(i)-p(i,j)*pq
            enddo 
         enddo
      enddo
      weight = w(1)
      v(1) = 1.d0
      ldv = 1
      call rstqlv (x,w,n,v,ldv,ierr)
      if (ierr .ne. 0) stop 'qrule 2'
      do j = 1,n
         w(j) = (weight*v(j))**2
      enddo
      return
      end

      subroutine rstqlv (d,e,n,v,ldv,ierr)
      implicit double precision (a-h,o-z)
c
c     ----------------------------------------------------------------- 
c     Eigenvalues and eigenvectors of a real symmetric tridiagonal
c     matrix. Based on the Numerical Recipes routine tqli but 
c     modified so as to calculate just the first ldv rows of the
c     eigenvector matrix. (Only the first row is required in the
c     discrete Stieltjes procedure.)
c
c     Note that v must be initialized to (the first ldv rows of)
c     a unit matrix before entry. 
c     ----------------------------------------------------------------- 
c
      dimension d(n),e(n),v(ldv,n)
c
      nv = min(ldv,n)
      do i = 2,n
         e(i-1) = e(i)
      enddo
      e(n) = 0.d0
      do l = 1,n
         iter = 0
   1     do m = l,n-1
            dd = abs(d(m))+abs(d(m+1))
            ee = abs(e(m))
            if (ee+dd .le. dd) goto 2
         enddo
         m = n
   2     if (m .ne. l) then
            if (iter .eq. 30) then
               ierr = 1
               return
            endif
            iter = iter+1
            g = (d(l+1)-d(l))/(2.d0*e(l))
            r = sqrt(1.d0+g**2)
            if (abs(g-r) .gt. abs(g+r)) r = -r
            g = d(m)-d(l)+e(l)/(g+r)
            s = 1.d0
            c = 1.d0
            p = 0.d0
            do i = m-1,l,-1
               f = s*e(i)
               b = c*e(i)
               r = sqrt(f**2+g**2)
               e(i+1) = r
               if (abs(r) .eq. 0.d0) then
                  d(i+1) = d(i+1)-p
                  e(m) = 0.d0
                  goto 1
               endif
               s = f/r
               c = g/r
               g = d(i+1)-p
               r = (d(i)-g)*s+2.d0*c*b
               p = s*r
               d(i+1) = g+p
               g = c*r-b
               do k = 1,nv
                  f = v(k,i+1)
                  v(k,i+1) = s*v(k,i)+c*f
                  v(k,i) = c*v(k,i)-s*f
               enddo
            enddo
            d(l) = d(l)-p
            e(l) = g
            e(m) = 0.d0
            goto 1
         endif
      enddo
      do j = 1,n-1
         k = j
         do i = j+1,n
            if (d(i) .lt. d(k)) k = i
         enddo
         if (k .ne. j) then
            swap = d(k)
            d(k) = d(j)
            d(j) = swap
            do i = 1,nv
               swap = v(i,k)
               v(i,k) = v(i,j)
               v(i,j) = swap
            enddo
         endif
      enddo
      ierr = 0
      return
      end

      subroutine rgfac (a,lda,n,indx,ierr)
      implicit double precision (a-h,o-z)
c
c     ----------------------------------------------------------------- 
c     LU decomposition routine for real matrices.
c
c     Uses the same pivoting strategy as Numerical Recipes ludcmp, 
c     but with a re-organization of the inner loops to exploit
c     sparsity and reduce the loop overhead.
c     ----------------------------------------------------------------- 
c
      dimension a(lda,n),indx(n)
      dimension vv(n)
c
      do i = 1,n
         aamax = 0.d0
         do j = 1,n
            if (abs(a(i,j)) .gt. aamax) aamax = abs(a(i,j))
         enddo
         if (aamax .eq. 0.d0) then
            ierr = 1
            return
         endif
         vv(i) = 1.d0/aamax
      enddo
      do j = 1,n
         do k = 1,j-1
            if (a(k,j) .ne. 0.d0) then
               do i = k+1,n
                  a(i,j) = a(i,j)-a(i,k)*a(k,j)
               enddo
            endif
         enddo
         aamax = 0.d0
         do i = j,n
            atest = vv(i)*abs(a(i,j))
            if (atest .ge. aamax) then
               imax = i
               aamax = atest
            endif
         enddo
         if (j .ne. imax) then
            do k = 1,n
               swap = a(imax,k)
               a(imax,k) = a(j,k)
               a(j,k) = swap
            enddo
            vv(imax) = vv(j)
         endif
         indx(j) = imax 
         if (a(j,j) .eq. 0.d0) then
            ierr = 2
            return
         endif
         if (j .ne. n) then
            pivot = 1.d0/a(j,j)
            do i = j+1,n
               a(i,j) = a(i,j)*pivot
            enddo
         endif
      enddo
      ierr = 0
      return
      end

      subroutine rgsol (a,lda,n,indx,b)
      implicit double precision (a-h,o-z)
c
c     ----------------------------------------------------------------- 
c     Uses the factorized A from rgfac to solve the linear 
c     equations A*X = B, overwriting the solution X on B.
c     Like the Numerical Recipes routine lubksb, but with
c     a slightly different calling sequence.
c     ----------------------------------------------------------------- 
c
      dimension a(lda,n),indx(n),b(n)
c
      ii = 0
      do i = 1,n
         ll = indx(i)
         sum = b(ll)
         b(ll) = b(i)
         if (ii .ne. 0) then
            do j = ii,i-1
               sum = sum-a(i,j)*b(j)
            enddo
         else if (sum .ne. 0.d0) then
            ii = i
         endif
         b(i) = sum
      enddo
      do i = n,1,-1
         do j = i+1,n
            b(i) = b(i)-a(i,j)*b(j)
         enddo
         b(i) = b(i)/a(i,i)
      enddo
      return
      end

\end{verbatim}

This program can be compiled with the gfortran compiler, using the -Os compiler option. It generates the following output file for Model I with $N=999$ and $M=3$: 

\begin{verbatim}

  N0 =     0   N =   999   M =     3

      j      N_j                 A_j

      1      278      0.006211535015
      2      444      0.027438527427
      3      277      0.048627307400

      k              % error in mu_k

      1                     0.000000
      2                     0.000000
      3                     0.000000

      4                     0.000005
 
\end{verbatim}

And the following output file for Model II with $N_0=24$ and $M=5$:

\begin{verbatim}

  N0 =    24   N =    48   M =     5

      j      N_j                 A_j

      1       12      0.045891672330
      2       16      0.088805260206
      3       11      0.161816389804
      4        6      0.231833880157
      5        3      0.281681731186

      k              % error in mu_k

      1                     0.000000
      2                     0.000000
      3                     0.000000
      4                     0.000000
      5                     0.000000

      6                     0.000567
      
\end{verbatim}

\section{Exploiting equivalent spins}

The number of ways $W(N,I)$ in which $N$ equivalent spin-1/2 nuclei can be combined to give a resultant spin with angular momentum quantum number $I$ is summarised in the following table:

\begin{table} [h]
\centering
\begin{tabular} { c r r r r r r r r r }
$N$ & \quad $I=0$ & \quad ${1\over 2}$ & \quad 1 & \quad ${3\over 2}$ & \quad 2 & \quad ${5\over 2}$ & \quad 3 & \quad ${7\over 2}$ &\quad 4 \\
\\
\hline
1 && 1 \\
2 & 1 && 1 \\
3 && 2 && 1 \\
4 & 2 && 3 && 1 \\
5 && 5 && 4 && 1 \\
6 & 5 && 9 && 5 && 1 \\
7 && 14 && 14 && 6 && 1 \\
8 & 14 && 28 && 20 && 7 && 1 \\
(etc.) \\
\hline
\end{tabular}
\end{table}

\noindent
The entries in this table are straightforward to generate on a computer using the recurrence relation
$$
W(N,I) = \begin{cases}
W(N-1,I+1/2), & I=0 \\
W(N-1,I-1/2)+W(N-1,I+1/2), & 0<I<N/2 \\
W(N-1,I-1/2), & I=N/2,
\end{cases}
$$
they are given explicitly by
$$
W(N,I) =\begin{pmatrix} N\\ N/2+I\end{pmatrix}{(2I+1)\over (N/2+I+1)},
$$
and one can show that they satisfy
$$
\sum_{I} W(N,I)(2I+1) = 2^N.
$$
I.e., the $2^N$ states in the uncoupled representation $\ket{\sigma_1,\ldots,\sigma_N}$ (where $\sigma_I=\pm1/2$ is the projection of the $i$-th nuclear spin on the $z$ axis) can be combined to give the same number of states $\ket{I,M_I}$ in the coupled representation, where $I$ ranges from mod$(N,2)/2$ to $N/2$ and $M_I$ ranges from $-I$ to $I$ in integer steps.  

It follows from the above table that a central spin problem with $N=4$ equivalent spin-1/2 nuclei can be reduced to 3 separate calculations, each of which involves a {\em single} nuclear spin with $I=0$, 1, or 2 coupled to the central electron spin. The results of these calculations are simply multiplied by the weight factors $W(N,I)=2$, 3, 1, and then added together and divided by $Z=2^{N+1}$ to obtain the central spin correlation tensor
$$
R_{\alpha\beta}(t) = {1\over Z}{\rm tr}\left[\hat{S}_{\alpha}(0)\hat{S}_{\beta}(t)\right].
$$
A central spin problem with one set of $N_1=4$ equivalent spin-1/2 nuclei and another set of $N_2=3$ equivalent spin-1/2 nuclei can be reduced in the same way to 6 separate calculations, each of which involves pair of nuclear spins with $(I_1,I_2)=(0,1/2)$, (0,3/2), (1,1/2), (1,3/2), (2,1/2), or (2,3/2) coupled to the central electron spin. The corresponding weights are now $W(N_1,I_1)W(N_2,I_2)=4$, 2, 6, 3, 2, 1, respectively. And so on. One can automate this procedure for an arbitrary number of sets of equivalent nuclei, and also generalise it to the case where the equivalent nuclei have spins other than 1/2.

\newpage
\section{Additional convergence tests}

Figure~1 of the paper shows the convergence of $R_{zz}(t)$ with increasing $M$ for Model I with $N=49$. The following figures show similar convergence tests for Model I with $N=99$ and 999, and for Model II with $N_0=24$ and 36.

\begin{figure}[h]
\centering
\resizebox{0.5\columnwidth}{!} {\includegraphics{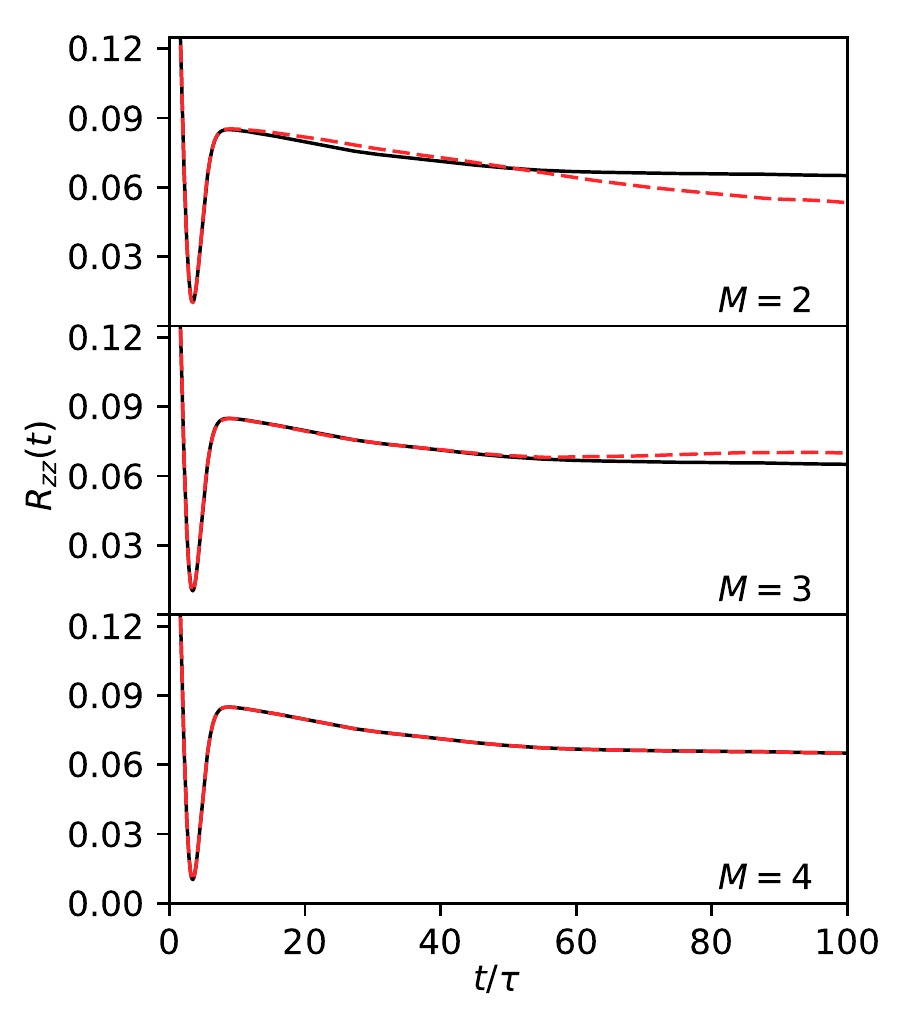}}
\caption{Convergence of $R_{zz}(t)$ for Model I with $N=99$, as a function of $M$. The solid black curve in each panel is the fully converged result obtained with $M=5$, and the dashed red curve is the result obtained with the specified value of $M$.}
\end{figure}

\begin{figure}[h]
\centering
\resizebox{0.5\columnwidth}{!} {\includegraphics{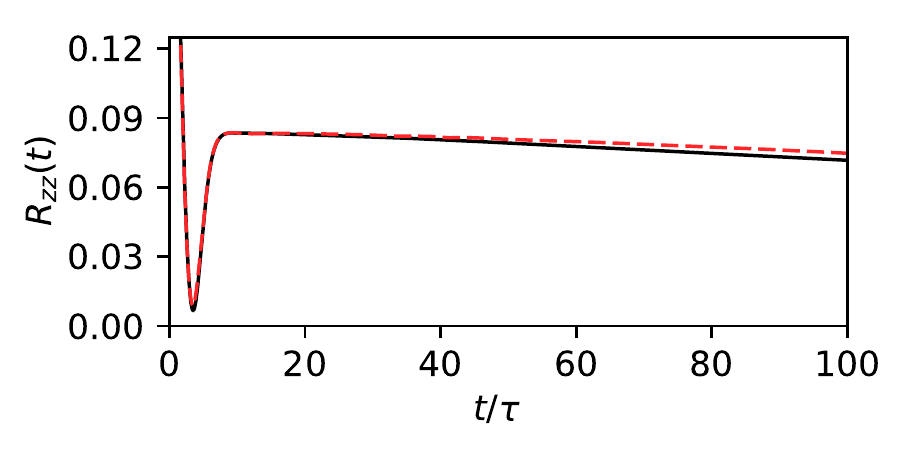}}
\caption{Comparison of the results obtained with $M=2$ (dashed red line) and $M=3$ (solid black curve) for $R_{zz}(t)$ of Model I with $N=999$. In this case, because of the large value of $N$, we did not have the computational resources to go up to $M=4$. However, it is already clear from this comparison and the other convergence tests we have presented that the $M=3$ results for $N=999$ are likely to be converged to graphical accuracy all the way out to $t=100\tau$.}
\end{figure}

\begin{figure}[h]
\centering
\resizebox{0.5\columnwidth}{!} {\includegraphics{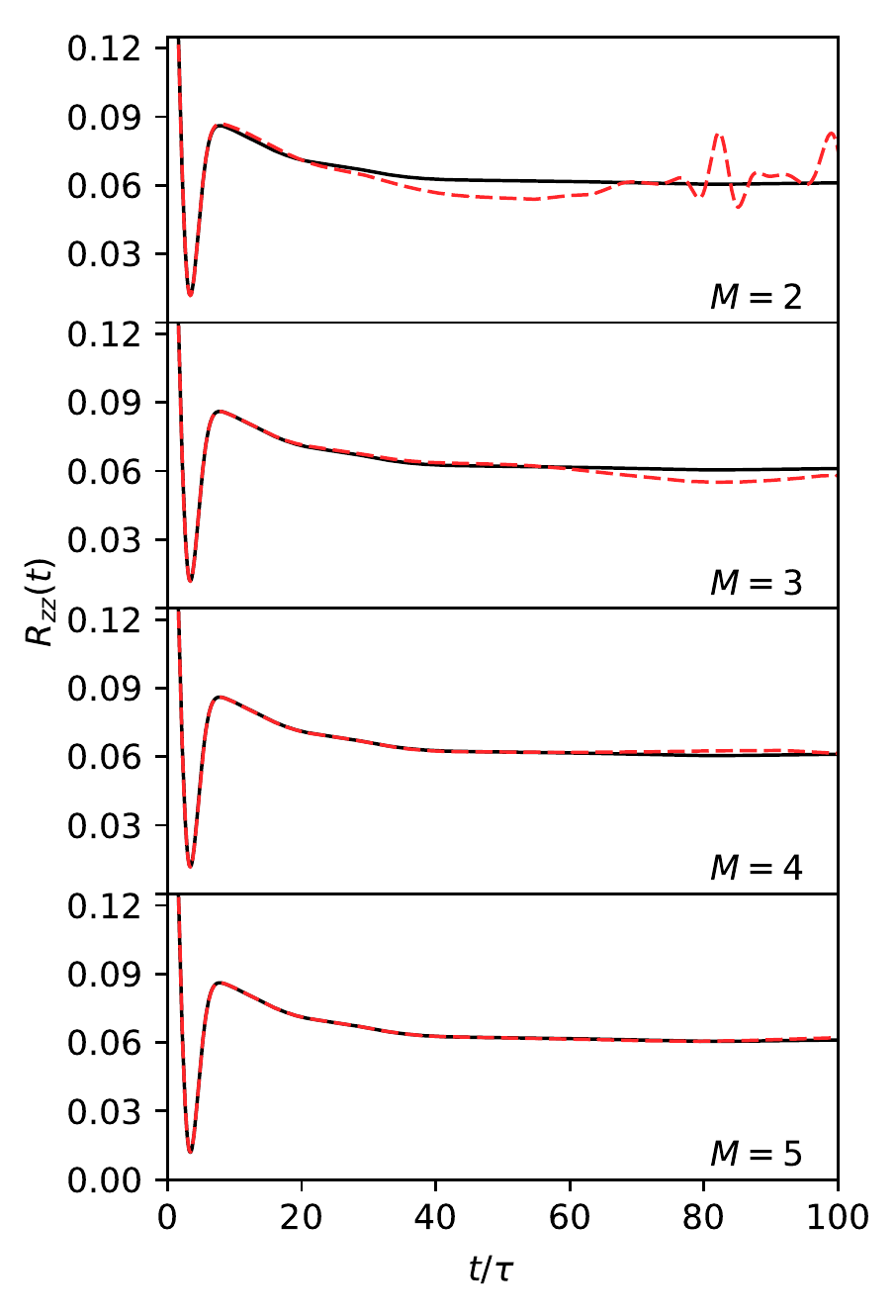}}
\caption{Convergence of $R_{zz}(t)$ for Model II with $N_0=24$, as a function of $M$. The solid black curve in each panel is the fully converged result obtained with $M=7$, and the dashed red curve is the result obtained with the specified value of $M$.}
\end{figure}

\begin{figure}[h]
\centering
\resizebox{0.5\columnwidth}{!} {\includegraphics{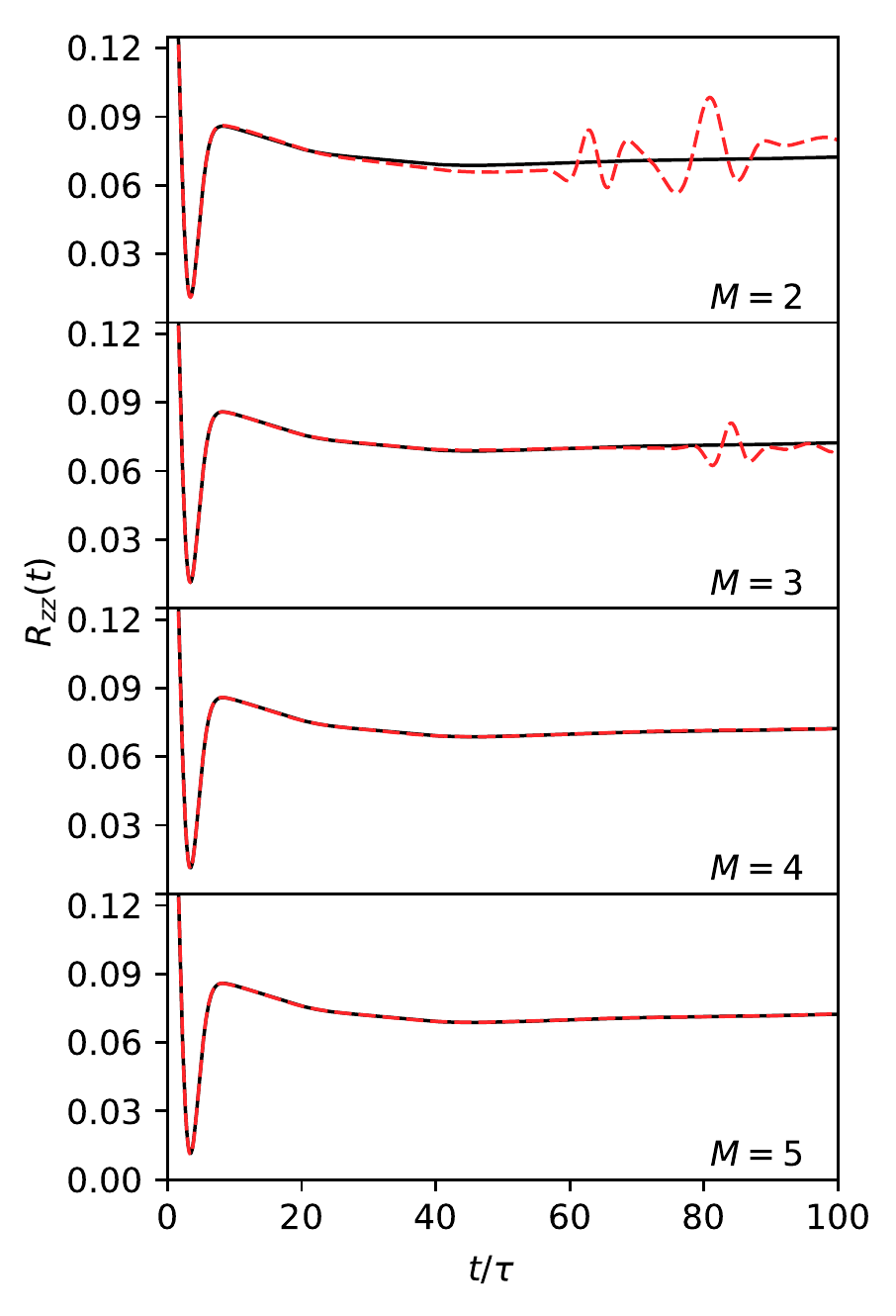}}
\caption{Convergence of $R_{zz}(t)$ for Model II with $N_0=36$, as a function of $M$. The solid black curve in each panel is the fully converged result obtained with $M=7$, and the dashed red curve is the result obtained with the specified value of $M$.}
\end{figure}

\newpage
\section{Long time correlations}

Figure~4 in the manuscript compares the Schulten-Wolynes (SW), improved semiclassical (SC), and quantum mechanical (QM) correlation functions $R_{zz}(t)$ for Model I with $N=49$, 99, and 999 nuclear spins. That figure was plotted with linear axes ($R_{zz}(t)$ versus $t$) to emphasise that the SW theory misses the long time decay of the central spin correlation function and the SC theory predicts too rapid a long time decay (see especially the insets in the $N=49$ and $N=99$ panels of the figure). 

The present Fig.~5 plots the same data on a semi-log plot ($R_{zz}(t)$ versus $\log t$). This makes it clearer that, at least for $N=49$ and 99, the SC and QM $R_{zz}(t)$'s have reached a plateau value by the time $t=100\tau$. If the (cheaper) SC calculation is extended to $t=200\tau$, the computed $R_{zz}(t)$ remains at this plateau value, and since the SC and QM calculations agree at $t=100\tau$ we expect that this would also be the case in the QM calculation. This suggests that, even with a finite number of nuclear spins in the central spin problem (here with a uniform distribution of hyperfine coupling constants, and in the absence of an applied magnetic field), the central spin retains some information about its initial state in the long time limit. This is especially relevant to the quantum dot problem because it is a prerequisite for being able to use a quantum dot as a qubit in a quantum computer (although of course in a real quantum dot the dipolar coupling between the nuclear spins -- which we have ignored in the present calculations -- will eventually play a role on a sufficiently long time scale).

The SC and QM results for $N=999$ in Fig.~5 have not yet reached their long-time limit at $t=100\tau$, but we suspect on the basis of the $N=49$ and 99 results that these $R_{zz}(t)$'s are also tending to a non-zero plateau value. We are not in a position to predict this value because the plateau values for $N=49$ and 99 are both the same ($0.0645$) to within the accuracy of our calculations ($\pm 0.001$). The plateau value for the SW theory, which corresponds to switching off the nuclear spin precession by taking the limit as $N\to\infty$ before the limit as $t\to\infty$, is $R_{zz}(t\to\infty)=1/12\sim 0.083$.

\begin{figure}[h]
\centering
\resizebox{0.5\columnwidth}{!} {\includegraphics{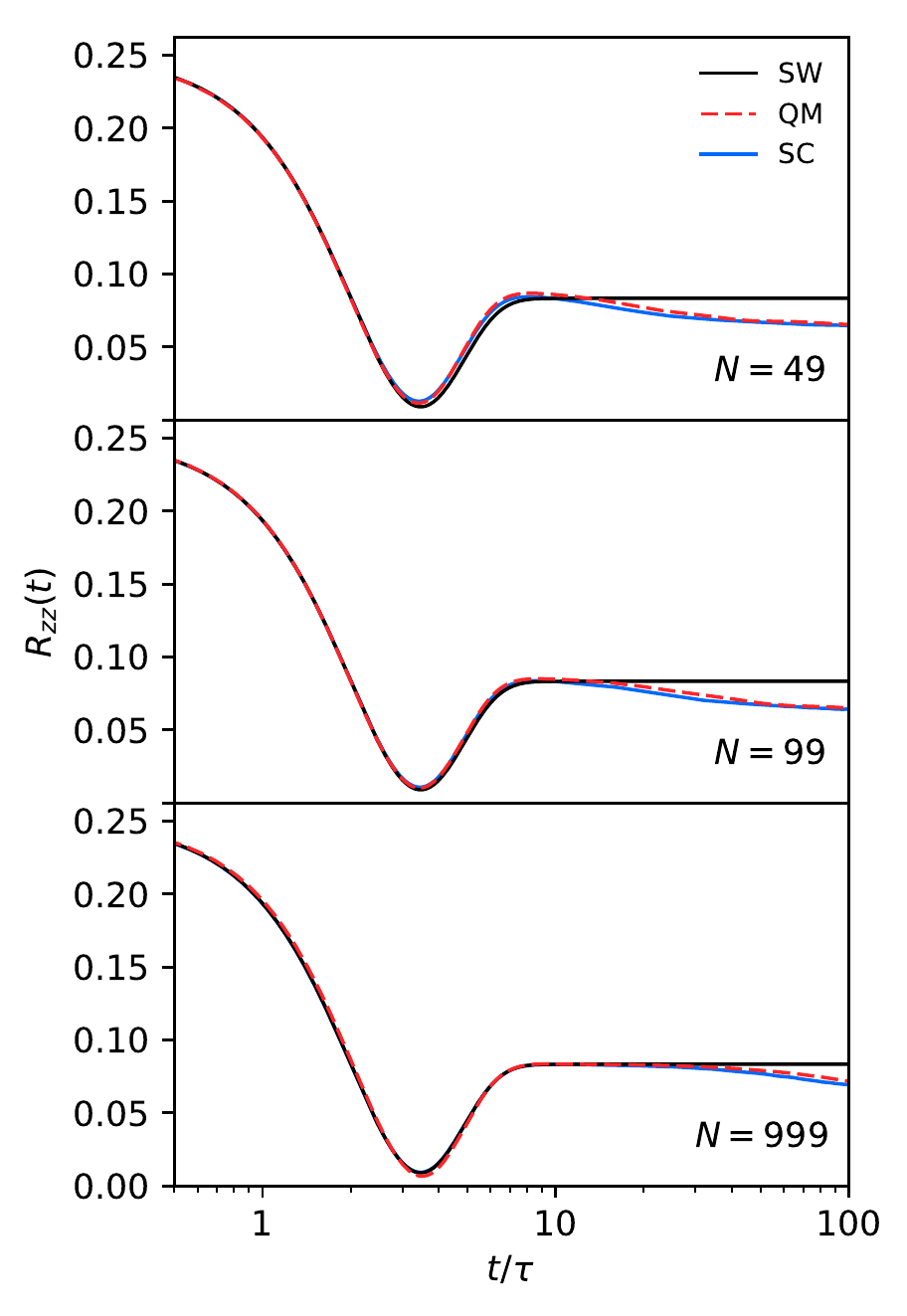}}
\caption{As in Fig.~4 of the paper, but with the data plotted on a semi-log plot ($R_{zz}(t)$ versus $\log t$) to emphasise the long-time behaviour of the Schulten-Wolynes (SW), semiclassical (SC) and quantum mechanical (QM) correlation functions.}
\end{figure}

It is also revealing to plot the data in Figure~5 of the manuscript on a semi-log plot ($R_{xx}(t)$ versus $t$) to bring out the long-time behaviour of the various (SW, SC, and QM) correlation functions. This plot is shown in the present Fig.~6. The SW theory is qualitatively wrong in this case -- $R_{xx}(t)$ with a finite magnetic field in the $z$ direction -- in predicting a finite plateau value in the long time limit. The SC and QM curves have not converged to their long time limits by the end of the plot ($t=400\tau$), but are both seen to be oscillating around zero. When we extend the (cheaper) SC calculation to longer times, we find that the amplitude of the oscillation decays to zero, and we would expect the same to be the case in the QM calculation. Combining this with the results for the other components of the spin correlation tensor (not shown in the figure), we find that a magnetic field of $1/4\tau$ in the $z$ direction causes complete decoherence of the central spin in the $xy$ plane ($R_{xx}(t\to\infty)=R_{xy}(t\to\infty)=R_{yy}(t\to\infty)=0$), but not in the $z$ direction ($R_{zz}(t\to\infty)\sim 0.0875\pm 0.001$).

\begin{figure}[h]
\centering
\resizebox{0.5\columnwidth}{!} {\includegraphics{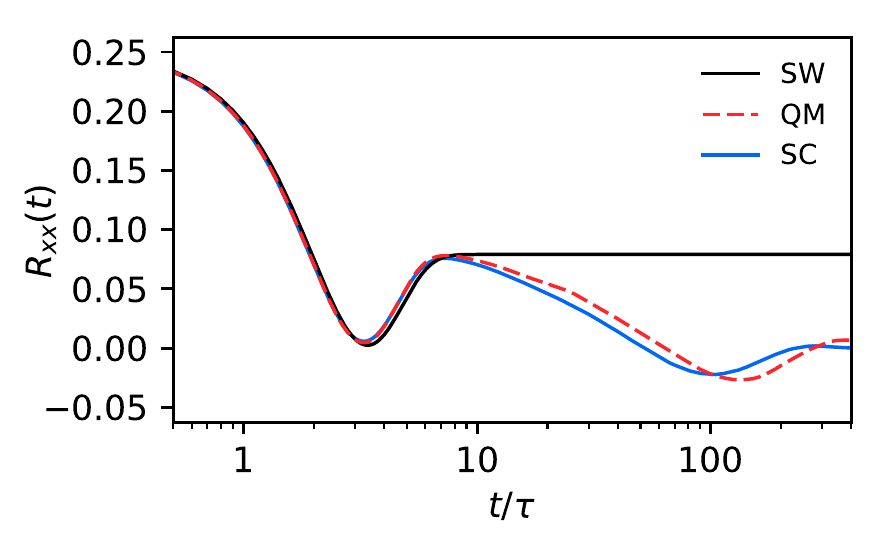}}
\caption{As in Fig.~5 of the paper, but with the data plotted on a semi-log plot ($R_{xx}(t)$ versus $\log t$) to emphasise the long-time behaviour of the Schulten-Wolynes (SW), semiclassical (SC) and quantum mechanical (QM) correlation functions..}
\end{figure}